\documentclass{aa} 
\def\apj{ApJ}%
\def\apjl{ApJ}%
\def\apjs{ApJS}%
\def\ao{Appl.~Opt.}%
\def\aap{A\&A}%
\usepackage{graphicx}
\usepackage{subfigure}  
\usepackage{txfonts}
\begin{document}
   \title{Rotational Structure and Outflow in the Infrared Dark Cloud 18223-3\thanks{The FITS files of the integrated continuum and line maps, as well as the first and second moment maps are available in electronic form at the CDS via anonymous ftp to cdsarc.u-strasbg.fr (130.79.128.5)
or via http://cdsweb.u-strasbg.fr/cgi-bin/qcat?J/A+A/}}


   \author{C. Fallscheer
          \inst{1}
          \and
          H. Beuther\inst{1}
	  \and
	Q. Zhang\inst{2}
	\and
	E. Keto\inst{2}
	\and
	T.K. Sridharan\inst{2}
          }

   \offprints{C. Fallscheer}

   \institute{Max Planck Institute for Astronomy, K\"{o}nigstuhl 17, 69117 Heidelberg Germany\\
              \email{fallscheer@mpia.de}
         \and
             Harvard-Smithsonian Center for Astrophysics, 60 Garden St., Cambridge, MA 02138\\
             }

   \date{Received April 8, 2009; accepted July 6, 2009}

 
  \abstract
   {}
   {We examine an Infrared Dark Cloud at high spatial resolution as a means to study rotation, outflow, and infall at the onset of massive star formation.} 
   {The Infrared Dark Cloud (IRDC)\ 18223-3 was observed at 1.1 mm and 1.3 mm with the Submillimeter Array (SMA) and follow-up short spacing information was obtained with the IRAM 30\,m telescope. 
  Additional data were taken at 3 mm with the IRAM Plateau de Bure Interferometer (PdBI).}
   {Submillimeter Array observations combined with IRAM 30 meter data in $^{12}$CO(2--1) reveal the outflow orientation in the IRDC\ 18223-3 region, and PdBI 3 mm observations confirm this orientation in other molecular species.  
The implication of the outflow's presence is that an accretion disk is feeding it, so using line data for high density tracers such as C$^{18}$O, N$_2$H$^+$, and CH$_3$OH, we looked for indications of a velocity gradient perpendicular to the outflow direction.  Surprisingly, this gradient turns out to be most apparent in
CH$_3$OH.  The large size (28,000 AU) of the flattened rotating object detected indicates that this velocity gradient cannot be due solely to a disk, but rather from inward spiraling gas within which a Keplerian disk likely exists. 
The rotational signatures can be modeled via rotationally infalling gas.  From the outflow parameters, we derive properties of the source such as an outflow dynamical age of $\sim$37,000 years, outflow mass of $\sim$13 M$_{\odot}$, and outflow energy of $\sim$1.7 $\times$ 10$^{46}$ erg.  While the outflow mass and energy are clearly consistent with a high-mass star forming region, the outflow dynamical age indicates a slightly more evolved evolutionary stage than previous spectral energy distribution (SED) modeling indicates.} 
   {The orientation of the molecular outflow associated with IRDC\ 18223-3 is in the northwest-southeast direction and velocity gradients orthogonal to the outflow reveal a large rotating structure likely harboring an accretion disk within.  
We also present a model of the observed methanol velocity gradient.  
The calculated outflow properties reveal that this is truly a massive star in the making. These data present evidence for one of the youngest known outflow/infall/disk systems in massive star formation.  A tentative evolutionary picture for massive disks is discussed.}

   \keywords{stars: formation --
                stars: individual (IRDC\ 18223-3) --
                stars: early type
               }
\authorrunning{C. Fallscheer et al.}
\titlerunning{Rotational Structure and Outflow in IRDC\ 18223-3}
   \maketitle
%

\section{Introduction}


Massive stars (M$>$8M$_{\odot}$) are paramount constituents of galaxies, yet their formation is significantly less well understood than low-mass star formation.
Theories of massive star formation predict massive accretion disks and indirect evidence for disks is prevalent, but observational evidence for such structures is rare.  Outflows appear to be widespread in massive star formation regions (\cite{she96}; \cite{zha01}, 2005\nocite{zha05}; \cite{beu02b}), and must be powered and fed by a (massive) accretion disk.  Evidence of rotation and infall that hint at the presence of disks has been detected in several systems, although further study is required (see reviews by \cite{beu07pp5}; \cite{ces07}; \cite{zin07}). 

Several factors complicate observational disk studies of massive star formation regions. Compared to low mass star formation regions, they are much rarer and further away on average.   Additionally, massive stars evolve quickly and typically form deeply embedded within a core at the center of a cluster, thus obscuring the process from optical wavelengths and making it difficult to distinguish one protostar from another in such a busy region of star formation typically at kpc distances.  Due to these complications, it is extremely difficult to directly observe massive accretion disks at sufficiently high spatial resolution to definitively identify the disk component.
Nevertheless, today's interferometers are conducive to studies that probe size-scales comparable to the sizes we might expect for accretion disks around massive stars, and studies of disks in massive star formation regions have recently increased significantly in number (\cite{ces07}). 

The detection of circumstellar disks would provide strong support for the theory that massive star formation is similar to a scaled-up version of low-mass star formation (e.g. \cite{yor02}; \cite{kru07}, 2009\nocite{kru09}).  These rotating disks, if present, would exhibit a velocity gradient perpendicular to the established outflow direction.  


While previous disk studies in high-mass star formation mainly concentrated on relatively evolved regions such as high-mass protostellar objects (e.g. IRAS 20126+4104, \cite{ces97}, 2005\nocite{ces05}; G192.16-3.82, \cite{she02}; IRAS 18089-1732, \cite{beu08}), here we focus on one of the earliest evolutionary stages, namely an Infrared Dark Cloud.

A survey of high-mass protostellar objects (HMPOs) associated with Infrared Astronomical Satellite (IRAS) sources was conducted (\cite{beu02a}), and 
IRDC\ 18223-3 happened to be in the field of view of one of the target sources, IRAS 18223-1243.  IRDC\ 18223-3 lies at a distance of 3.7 kpc (\cite{sri05}) in a filamentary structure extending south from IRAS 18223-1243 (\cite{beu05b}).
IRDC\ 18223-3 is likely a high-mass core containing a young stellar object that is currently accreting at a high enough rate that it will eventually join the high-mass regime (\cite{beu07}).  Our current studies present the outflow properties of this source and reveal velocity gradients in several molecular species that are not aligned with the outflow orientation.  

Based on the analysis of Spitzer observations combined with SMA observations, IRDC\ 18223-3 is comprised of at least two separate components (\cite{beu07}).  A cooler component at $\sim$15 K contains most of the mass (580 M$_{\odot}$) while a warmer component at $\sim$50 K with approximately 0.01 M$_{\odot}$ is necessary to explain the excess flux in the SED at 24 $\mu$m.  The young source has an accretion-dominated
 luminosity of 180 L$_{\odot}$ that is expected to increase over time.  While the young stellar object has not yet accumulated a significant amount of mass, in the framework of the simulations by Krumholz et al. (2007)\nocite{kru07}, the source's mass and luminosity are consistent with a large accretion rate of $\sim$10$^{-4}$ M$_{\odot}$ yr$^{-1}$.  This high accretion rate, as well as the outflow parameters we derive, both point to a scenario that
this source will eventually become a massive star 
and that we are looking at one of the earliest stages of massive star formation.
Therefore, this source is an excellent case study for looking at the properties of massive stars at the beginning stages of their lifetimes.  This source also provides an ideal specimen to study the role of disks in early evolutionary phases of massive star formation.

In this paper we report a cone-shaped structure detected in $^{12}$CO(2--1) outlining the molecular outflow as well as a velocity gradient of 3 km s$^{-1}$ in CH$_3$OH(6$_K$--5$_K$) and N$_2$H$^+$(3--2) perpendicular to the outflow.  We also see velocity gradients on the order of a few km s$^{-1}$ not aligned with the outflow axis in C$^{18}$O(2--1). 
Additionally, we present the results of modeling the observed CH$_3$OH parameters to obtain the velocity and density structure of the region.

\section{Observations}

\begin{table}
\caption{Observed lines}             
\label{obs_lines}      
\centering                          
\begin{tabular}{llrr}        
\hline\hline                 
Telescope & Transition 			& Rest Freq.    & E$_{upper}$ 	\\    
	  &				& [GHz] 	& [K]	\\
\hline
PdBI     & H$^{13}$CO$^{+}$ 1 $\rightarrow$ 0 	& 86.754	& 4	\\
PdBI     & SiO 2 $\rightarrow$ 1			& 86.847	& 6	\\
PdBI     & HN$^{13}$C 1 $\rightarrow$ 0		& 87.091	& 4	\\
SMA      & C$^{18}$O 2 $\rightarrow$ 1		& 219.560	& 16	\\
SMA      & $^{13}$CO 2 $\rightarrow$ 1		& 220.399	& 16	\\
SMA      & CH$_3$OH 8$_{-1}$ $\rightarrow$ 7$_0$ E	& 229.759	& 89	\\
SMA, 30m & CO 2 $\rightarrow$ 1			& 230.538	& 17	\\
SMA      & N$_2$H$^+$ 3 $\rightarrow$ 2			& 279.512	& 27	\\
SMA      & CH$_3$OH 6$_0$ $\rightarrow$ 5$_0$ E	& 289.939	& 62	\\
SMA      & CH$_3$OH 6$_{-1}$ $\rightarrow$ 5$_{-1}$ E & 290.070	& 54	\\
SMA      & CH$_3$OH 6$_{0}$ $\rightarrow$ 5$_{0}$ A	& 290.111	& 48	\\
SMA      & CH$_3$OH 6$_{1}$ $\rightarrow$ 5$_{1}$ E	& 290.249	& 70	\\
SMA      & CH$_3$OH 6$_{-2}$ $\rightarrow$ 5$_{-2}$ E & 290.307	& 75	\\
\hline
\end{tabular}
\end{table}

Observations with the Submillimeter Array, Plateau de Bure Interferometer and IRAM 30\,m observatories include the 1.3 mm dust continuum and several spectral lines (see Table \ref{obs_lines}). The details of the observations are discussed below.

\subsection{Submillimeter Array}

Observations of IRDC 18223-3 with the Submillimeter Array$\footnote{The Submillimeter Array is a joint project between the Smithsonian Astrophysical Observatory and the Academia Sinica Institute of Astronomy and Astrophysics, and is funded by the Smithsonian Institution and the Academia Sinica.}$ were carried
out the nights of 30 May 2006, 8 August 2006, 24 July 2007, and 27 August 2007.  The phase reference center of this source is R.A. 18h25m08.55s Decl.=-12$^{\circ}$45$'$23.3$''$
(J2000.0) and the velocity of rest $v_{lsr}$ is 45.3 km s$^{-1}$ (\cite{sri05}).  The SMA has two spectral sidebands--both 2 GHz wide separated by 10 GHz.  All SMA data sets were calibrated with the IDL superset MIR$\footnote{The MIR cookbook by Charlie Qi can be found at http://cfa-www.harvard.edu/~cqi/mircook.html}$, then imaged and analyzed with MIRIAD.  These submillimeter interferometric observations allow us to probe through the optically opaque molecular cloud which IRDC 18223-3 is deeply embedded within.  Specific details of the observations are listed in Table~\ref{obs_params}.

The observations on 30 May 2006 are with all 8 antennas at 1.3 mm in the extended configuration. 
Zenith opacity measured by the Caltech Submillimeter Observatory was approximately $\tau$(230) 0.15.   Observing with 8 antennas coupled with the fact that
the track was 8.5 hours long (5 hours on-source integration) provided optimal coverage in the uv plane.  Passband and flux calibrations were derived from observations of Uranus.  For the flux, solutions could not be found on all baselines so it was necessary to perform an initial flux calibration on 3C279 to set the relative flux levels.  Phase and amplitude calibrations were made from regular observations (5 minutes every half hour) of quasars 1911-201 and 1733-130.
The same spectral setup was observed on 8 August 2006 in the compact configuration.  Only 7 antennas were used in this data set.  Despite $\tau$ varying between 0.2 and 0.25, the data are good quality. 
The same calibration objects were used as in the 30 May data set, however, it was not necessary to set the relative flux levels before using Uranus for the flux calibration.  The track was 7 hours long, 4 hours of which were spent on-source.  After combination of the data from both configurations, projected baselines range from 9 to 163 k$\lambda$.
 
Observations at 1.1 mm were obtained on 24 July 2007 in the compact configuration with all 8 antennas and in the extended configuration with only 6 fully--functioning antennas on 27 August 2007.  Zenith opacity, $\tau$(230 GHz), was $\sim$0.15 for the compact configuration observations, and $\sim$0.17 for the extended configuration observations. 
The quasar 1911-201 was used for phase and amplitude calibrations for the compact configuration data, and quasar 1733-130 was used for the extended configuration data.  Uranus was used for flux calibration in both 1.1 mm data sets and for phase and amplitude calibration of the compact configuration.  Quasar 3C454.3 was used for bandpass calibration of the extended configuration data as well as to set relative flux levels prior to using Uranus for the absolute flux calibration.  
Projected baselines ranged from 11 to 189 k$\lambda$ in the combined 1.1 mm data set.

\begin{table*}
\caption{Plateau de Bure Interferometer (PdBI), Submillimeter Array (SMA), and IRAM 30\,m observation parameters.}             
\label{obs_params}      
\centering                          
\begin{tabular}{c|c|c|c|c c c|c|c|c}        
\hline\hline                 
Obs & Date & Freq & Configuration 	& \multicolumn{3}{c|}{Calibrators} & Beam & rms$_{cont}$ & rms$_{line}$ \\    
&      & [GHz]  & 			& Bandpass & Phase \& Amplitude & Flux 	& [$''$] 	& [mJy/bm] & [mJy/bm]\\
\hline
& 23 May 2006 &  89 & D	     & 3C273   & NRAO 530 \& 1741-038 & MWC 349 &     	  &     & \\ 
PdBI & 03 Apr 2007 &  89 & C & 3C84    & NRAO 530 \& 1741-038 & MWC 349 &         &     & \\ 
& 05 Apr 2007 &  89 & C	     & 3C273   & NRAO 530 \& 1741-038 & MWC 349 &         &     & \\ 
& 	    &  89 & combined & 	       &		      &		& 6.7$\times$3.1 & 0.15 & 11\\
\hline                        
& 30 May 2006 & 230 & extended & Uranus  & 1911-201 \& 1733-130 & Uranus & & &\\ 
& 08 Aug 2006 & 230 & compact  & Uranus  & 1911-201 \& 1733-130 & Uranus & & &\\ 
SMA & 	    & 230 & combined &	       &		      &	       & 1.4$\times$1.3  & 1.6	& 65\\ 
& 24 Jul 2007 & 279 & compact  & Uranus  & 1911-201             & Uranus & & &\\ 
& 27 Aug 2007 & 279 & extended & 3C454.3 & 1733-130             & Uranus & & &\\ 
& 	    & 279 & combined & 	       &		      &	       & 1.3$\times$1.3  & 2.2	& 85\\ 
\hline
30\,m & 24 Feb 2007 & 230 & \multicolumn{4}{c|}{on-the-fly mapping} & 11 & & 0.17$^a$\\ 
\hline
\end{tabular}
\footnotesize{
Entries include the observatory at which the data were taken, the date of observation, the frequency observed, the array configuration, the calibrators, the synthesized beam obtained after inverting and cleaning the data, the 1$\sigma$ continuum noise level, and the 1 km s$^{-1}$ rms.\\
$^a$Units: K at 0.4 km s$^{-1}$ resolution averaged over the spectra in the central 4$\arcsec\times$4$\arcsec$ of the map.}
\end{table*}

\subsection{Plateau de Bure Interferometer}
Additional observations of IRDC 18223-3 were obtained at 3mm with the IRAM Plateau de Bure
Interferometer (PdBI)$\footnote{IRAM is supported by INSU/CNRS (France), MPG (Germany) and IGN (Spain)}$ located near Grenoble, France. Observations were carried out on 23 May 2006, then 3 and 5 April 2007.  These observations include detections of SiO(2-1), SO$_{2}$, HN$^{13}$C, and H$^{13}$CO$^+$(1-0) as well as the 3 mm continuum.  Quasars NRAO 530 and 1741-038 were used as phase and amplitude calibrators and MWC 349 for flux calibration.  Further details are listed in Table~\ref{obs_params}.

\subsection{Pico Veleta 30\,m Telescope}

To complement our interferometric data, we obtained short spacing observations of IRDC 18223-3 on 24 February 2007 with the HERA receiver at the IRAM 30\,m telescope near Granada, Spain. The $^{12}$CO(2--1) line at 230.5 GHz and $^{13}$CO(2--1) at 220.4 GHz
were observed in on-the-fly mode, resulting in maps approximately 140$''$ $\times$ 140$''$ in extent.  The sampling interval was 2$''$ and the region was scanned four times in the north-south direction and four times in the east-west direction in order to reduce effects caused by the scanning process.  The spectra were reduced using CLASS which is part of the GILDAS software package and then right ascension and declination scans were combined with the plait algorithm in GREG, another component of the GILDAS collection.  The $^{12}$CO data have a beam size of 11$\arcsec$ and a T$_{mb}$ corrected rms noise level of 0.17 K at 0.4 km s$^{-1}$ resolution (averaged over the spectra in the central 4$\arcsec\times$4$\arcsec$ region of the map).




After reducing the 30\,m data set separately, single dish $^{12}$CO data were converted to visibilities and then combined with the SMA data sets using the MIRIAD package task UVMODEL.  The synthesized beam of the combined data is 1.9$\arcsec$ x 1.7$\arcsec$ (PA 70$^{\circ}$).

\section{Results and Discussion}

\begin{figure}
\includegraphics[angle=-90,scale=0.6]{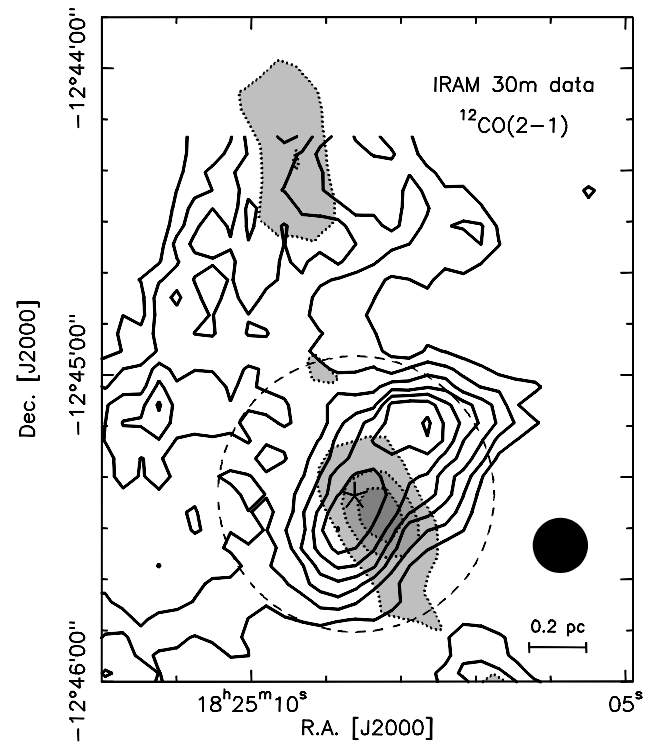}
\caption{Image of the velocity integrated $^{12}$CO(2--1) flux from the IRAM 30m single dish telescope of IRDC\ 18223-3. The integration is from 30-60 km\ s$^{-1}$ which includes the entire velocity range of the line.  Dotted contours of the 1.2 mm single dish continuum map (\cite{beu02a}) are overlaid and shaded in grayscale.  The star marks the position of the 1.3 millimeter continuum peak which is approximately 3.5$\arcmin$ south of IRAS 18223-1243. We attribute the difference in peak position between the 1.2 mm single dish and 1.3 mm interferometer data to pointing error in the single dish data.  
The size of the SMA primary beam (54$\arcsec$) is marked by the dashed circle.  The beam size is shown in the bottom right corner. \label{30m}}
\end{figure}

\begin{figure*}
\includegraphics[angle=-90,width=\textwidth]{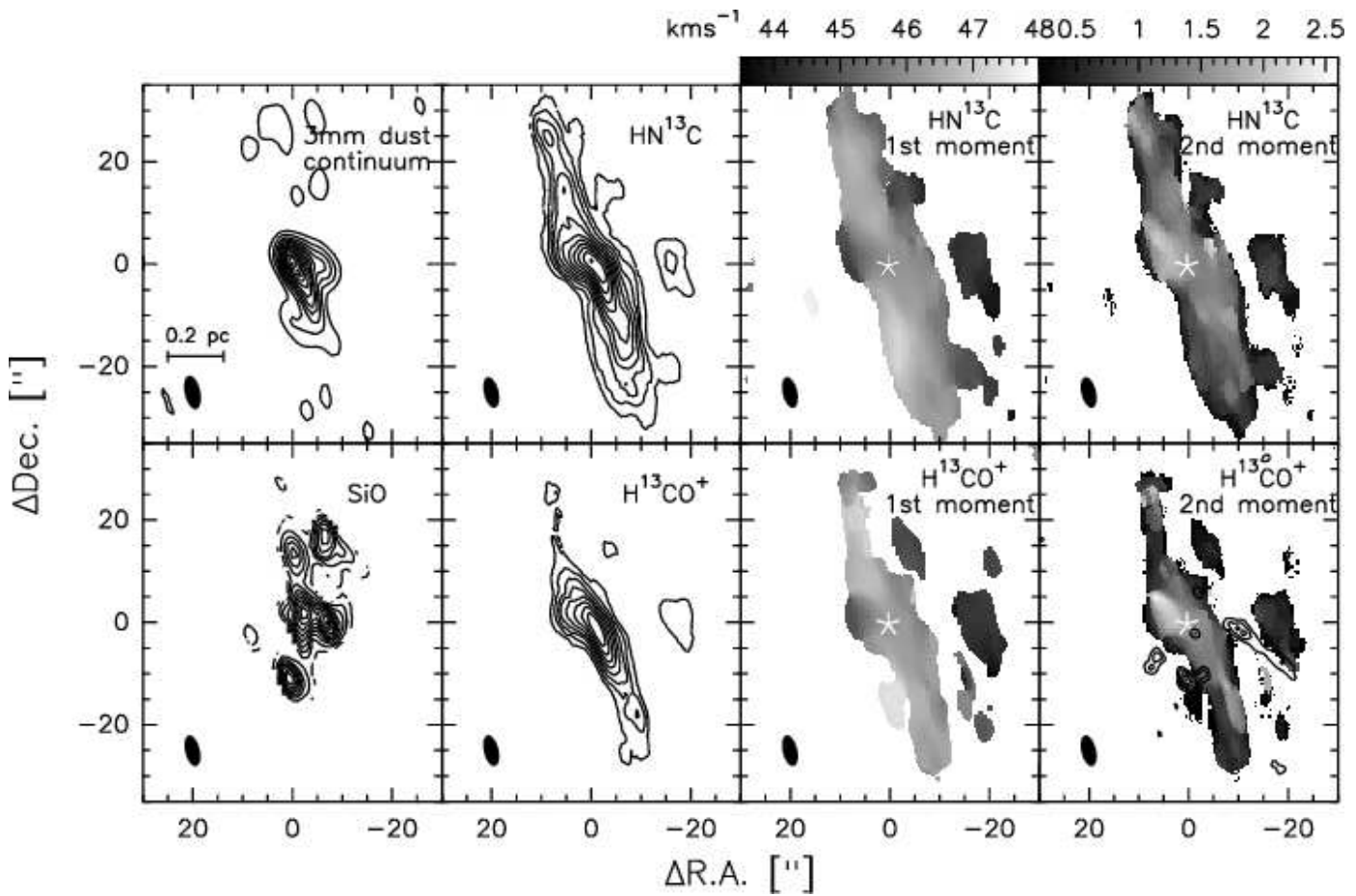}
\caption{Integrated flux of IRDC\ 18223-3 observed with PdBI of the 3mm dust continuum, SiO, HN$^{13}$C, and H$^{13}$CO$^+$. The integrations include the entire velocity range of the lines, namely 35-54 km\ s$^{-1}$ for SiO, and 43-48 km\ s$^{-1}$ for both HN$^{13}$C and H$^{13}$CO$^+$.  Contours always start at the 3$\sigma$ level.  For the continuum, SiO, HN$^{13}$C, and H$^{13}$CO$^+$ respectively, contours increase in steps of 3, 5, 3,and 8 $\sigma$ where $\sigma$ is 0.15, 11, 16, and 11 mJy\ bm$^{-1}$.  The velocity (1st) and line width (2nd) moment maps of HN$^{13}$C and H$^{13}$CO$^+$ are presented in the panels on the right.  The redshifted CO emission contours are overlaid on the H$^{13}$CO$^+$ 2nd moment map.  The star indicates the position of the dust continuum peak. 
The size of the synthesized beam is shown by the ellipse at the bottom left corner of each panel. \label{pdbi}}
\end{figure*}

\subsection{Large Scale Emission}

IRDC 18223-3 is about 3.5$\arcmin$ south of IRAS 18223-1243 in a filamentary structure extending more than 5$\arcmin$ ($>$5 pc) roughly in the north-south direction across the sky. In the large scale map presented in Figure \ref{30m}, the 1.2 mm continuum from previous observations (\cite{beu02a}) highlights the filamentary structure, and the $^{12}$CO(2--1) mostly traces emission associated with the outflow.  We also detect a weaker peak approximately 40$''$ east and slightly north of the primary peak.  However, as it is outside the primary beam of the interferometer observations, we do not detect this peak in subsequent observations discussed here.  It is interesting to note that this CO(2--1) peak is not associated with any emission in the 1.2 mm dust continuum.
 

Figure \ref{pdbi} presents the PdBI observations of several molecular species along with the dust continuum.  We defer discussion of SiO, a shock tracer, to section \ref{outflowprops}, Outflow Properties, because separate blueshifted and redshifted integrated maps (Figure \ref{outflow_sio}) indicate that it is associated with outflow activity.  HN$^{13}$C and H$^{13}$CO$^+$ both trace the large filamentary structure associated with IRAS 18223-1243 seen previously by Beuther et al. (2005c)\nocite{beu05c}.  Even the 3 mm dust continuum 
exhibits elongation along the filamentary structure.  The slight protrusion westward of the central 3 mm continuum peak may come about as a contribution from the outflow as there is also a peak in SiO at approximately the same spatial position.  As we do not see any indication of large scale velocity gradients in the first (velocity) moment maps of HN$^{13}$C and H$^{13}$CO$^+$, there likely is no significant movement happening along the filament although we cannot exclude the expansion of gas in the plane of the sky.  In contrast to previous N$_2$H$^+$(1--0) data (\cite{beu05b}), we do not see an increase in line width directly centered on IRDC 18223-3 in the second (line width) moment maps. 
Rather, 
we see increased line width to the northeast of the central peak.  The position of this peak in the second moment maps does not appear to be related to the outflow.

\subsection{Millimeter Continuum Emission}\label{mmcontinuum}

\begin{figure*}
\begin{center}
\includegraphics[angle=-90,width=\textwidth]{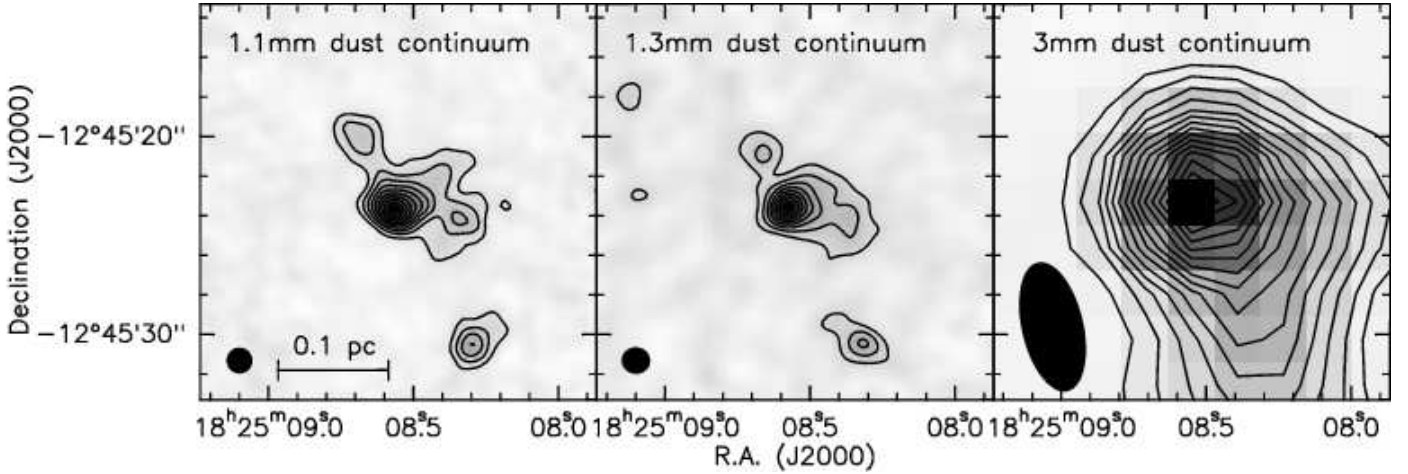}
\caption{The 1.1 mm (SMA), 1.3 mm (SMA), and 3.4 mm (PdBI) dust continuum.  The area displayed for all three panels is identical.  Contours start at 3$\sigma$ and increase in steps of 3$\sigma$ where $\sigma$ is 2.2, 1.6, and 0.15 mJy\ bm$^{-1}$ for the 1.1, 1.3, and 3.4 mm continuum images respectively.  } 
\label{cont}
\end{center}
\end{figure*}

The interferometer continuum data (Figure \ref{cont}) resolve the central structure into several peaks at a spatial resolution of $\sim$4000 AU.
The 1.1 mm dust continuum traces one main peak with 65 mJy bm$^{-1}$ and three secondary peaks.  A secondary peak of 17 mJy bm$^{-1}$ occurs 4.0$''$ to the northeast (P.A.=26$^{\circ}$) and another peak of 22 mJy bm$^{-1}$ is situated 3.2$''$ to the southwest (P.A.=261$^{\circ}$)
These secondary peaks as well as a 27 mJy bm$^{-1}$ peak that is more removed from the main peak (7.9$''$ with P.A.=209$^{\circ}$) 
are also detected at 1.3 mm, but an extension to the northwest in the 1.1 mm continuum is not seen at 1.3 mm.  While the secondary peaks may be separate sub-sources, the line data presented below indicate that the  two near-by peaks to the northeast and southwest are likely part of a larger scale infalling and rotating structure.  The spatial resolution of the 3 mm data is worse, but even in the zoomed in continuum image of Figure \ref{cont}, the extension tracing the filament toward the south is already evident.


With the flux measured within the outer contour of the 1.3 mm dust continuum, we estimate the gas mass M and beam averaged column density $N_{H_2}$ of the region following the methods of Hildebrand (1983) \nocite{hil83} and Beuther et al. (2002a, 2005b) \nocite{beu02a}\nocite{beu05b}.  These estimates are under the assumptions that the dust continuum is optically thin at this wavelength, and that the gas to dust mass ratio is 100.  For these calculations we assume a temperature of 20 K and a dust opacity index $\beta$ of 2 which corresponds to an opacity of 0.35 cm$^2$g$^{-1}$ at 1.3 mm.  Although Beuther \& Steinacker (2007) \nocite{beu07} determine components at two different temperatures, the 20 K we assume here is closer to
the temperature of the component with the majority of the mass in IRDC 18223-3.  We measure a total integrated flux of 130 mJy and a peak flux of 41 mJy bm$^{-1}$.  From these values we derive a total mass of 120 M$_{\odot}$ and a beam averaged column density of 3.8 x 10$^{24}$ cm$^{-2}$ corresponding to a peak visual extinction, A$_{\nu}$(mag), of $\sim$4000, where A$_{\nu}$ = N$_{H_{2}}$/0.94 $\times$ 10$^{21}$ cm$^{-2}$ mag$^{-1}$ (\cite{fre82}).  Taking the uncertainties of the temperature and dust opacity into account, we estimate that the calculated mass and beam averaged column density for the compact structures detected here are accurate to within a factor of approximately 5. 

It must also be noted that the mass estimate is heavily affected by missing flux in the interferometer data. Comparing the flux from the 250 GHz single dish bolometer map (dotted contours in Figure \ref{30m}) with the 230 GHz SMA continuum flux, we estimate that without short spacing information, the interferometer filters out about 80\% of the flux.  Hence we recover only about 20\% of the mass in our estimate.  
As a detection limit, we calculate a mass of 5.2 M$_{\odot}$ and a column density of 5.3 x 10$^{23}$ cm$^{-2}$ at the 3$\sigma$ (5.7 mJy bm$^{-1}$) flux level.  The interferometer data are not sensitive to structures less massive than this.  

\subsection{Outflow Properties} \label{outflowprops}

\begin{figure*}
\begin{center}
 \includegraphics[angle=-90,width=\textwidth]{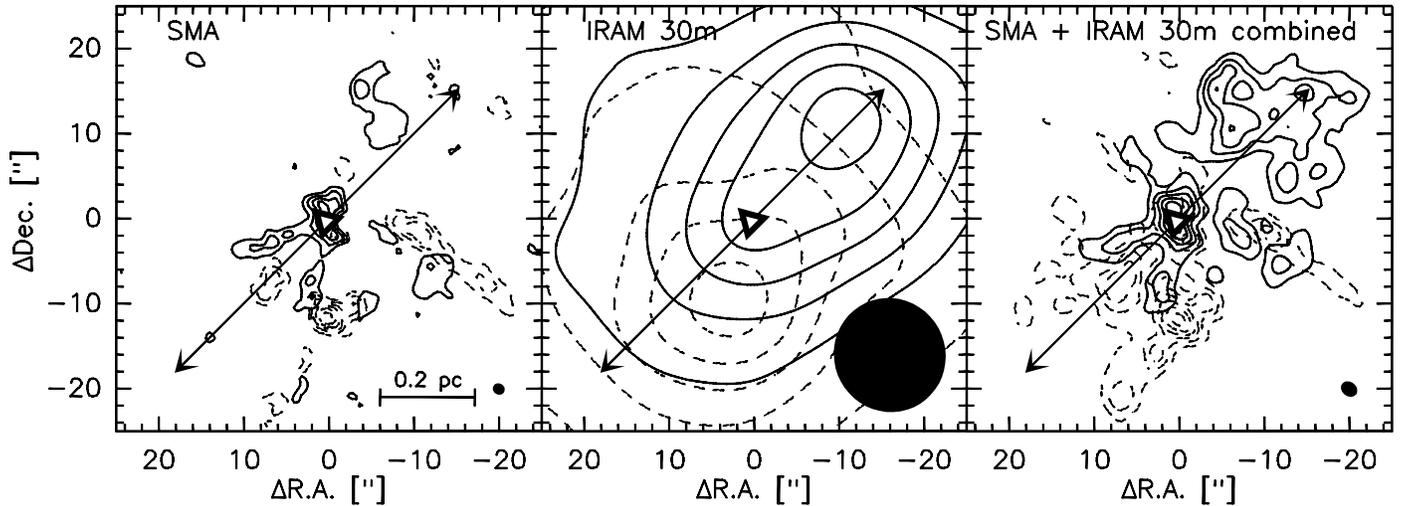}
 \caption{$^{12}$CO(2--1) integrated maps of IRDC 18223-3 from SMA, IRAM 30\,m, and SMA + IRAM 30\,m combined in the left, center, and right panels respectively.  Panels that include SMA data contain both extended and compact configuration observations.  We define the outflow orientation (indicated by the arrows) by the blueshifted interferometer component.  In each panel, the 1.3 mm continuum peak is indicated by the triangle, and the beam is included in the lower right corner. \textbf{Left:}  Blueshifted emission (solid contours) integrated over velocities of 38.5-43.5 km s$^{-1}$ in steps of 3$\sigma$ starting at 3$\sigma$ where $\sigma$=0.12 Jy bm$^{-1}$.  Redshifted emission (dashed contours) integrated over velocities of 50-54 km s$^{-1}$ in steps of 2$\sigma$ starting at 3$\sigma$ where $\sigma$=0.17 Jy bm$^{-1}$. \textbf{Center:} Blueshifted emission (solid contours) integrated over velocities of 36.5-43.5 km s$^{-1}$ in steps of 1$\sigma$ starting at 1$\sigma$ where $\sigma$=6.4 Jy bm$^{-1}$.  Redshifted emission (dashed contours) integrated over velocities of 46.5-53.5 km s$^{-1}$ in steps of 1$\sigma$ starting at 1$\sigma$ where $\sigma$=6.9 Jy bm$^{-1}$. \textbf{Right:} Blueshifted emission (solid contours) integrated over velocities of 36-44 km s$^{-1}$ in steps of 6$\sigma$ starting at 3$\sigma$ where $\sigma$=0.14 Jy bm$^{-1}$.  Redshifted emission (dashed contours) integrated over velocities of 50-56 km s$^{-1}$ in steps of 6$\sigma$ starting at 3$\sigma$ where $\sigma$=0.14 Jy bm$^{-1}$.
 \label{outflow}}
 \end{center}
 \end{figure*}

Figure 
\ref{outflow} demonstrates that there is a large scale outflow in the northwest-southeast orientation.  
The morphology of the blueshifted CO component to the southeast is cone-shaped and likely is tracing the outer walls of the outflow cavity.  There is also large, not very collimated blueshifted emission to the northwest of the source seen predominantly in the 30\,m data.  Although much less prevalent, we detect some redshifted emission on both sides of the primary 1.3 mm continuum peak.  The fact that we detect red- and blueshifted emission on both sides of the driving source indicates that the outflow is roughly in the plane of the sky because as the outflow cavity widens, the near edge would appear blueshifted while the far edge would appear redshifted.  



\begin{figure}
\begin{center}

\includegraphics[angle=-90,scale=.4]{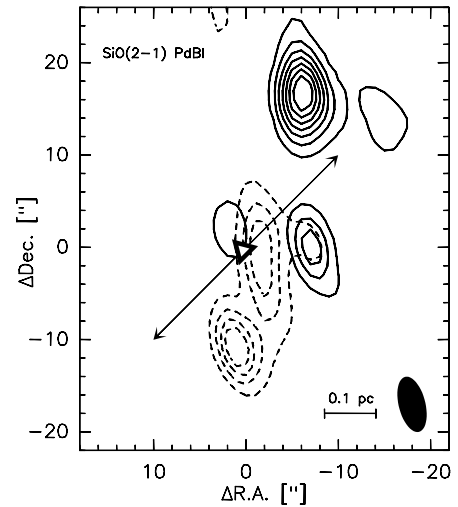}
 \caption{The SiO(2-1) integrated map.  Again the triangle symbol marks the position of the dust continuum peak.  The arrows indicate the orientation of the $^{12}$CO(2-1) outflow as shown in Figure \ref{outflow}.  The PdBI beam is included in the lower right corner.  Blueshifted emission (solid contours) integrated over velocities of 35-39 km s$^{-1}$ in steps of 4$\sigma$ starting at 3$\sigma$ where $\sigma$=0.011 Jy bm$^{-1}$.  Redshifted emission (dashed contours) integrated over velocities of 50-54 km s$^{-1}$ in steps of 4$\sigma$ starting at 3$\sigma$ where $\sigma$=0.011 Jy bm$^{-1}$.}  \label{outflow_sio}
 \end{center}
 \end{figure}

We detect SiO toward IRDC\ 18223-3 with the PdBI (See Figures \ref{pdbi} and \ref{outflow_sio}).  This tracer of shocked material, especially in molecular outflows (\cite{sch97}), is consistent with the outflow orientation we determine from $^{12}$CO(2--1) observations, and the southeastern SiO component coincides quite well with the southern edge of the blueshifted $^{12}$CO cone-shaped emission.  The slight difference in alignment between blue- and redshifted SiO emission shown in Figure \ref{outflow_sio} and the CO outflow orientation indicated by the arrows overlaid in this figure could potentially arise from the presence of multiple outflows. Such a scenario would indicate the presence of an undetected secondary source.  We discuss the possibility of multiplicity further in section \ref{modeling}.  Alternatively, the difference in morphology between SiO and CO may be explained by precession of the jet/outflow in which the SiO emission traces the current jet component's location.  The fact that SiO tends 
to trace the more collimated jet component whereas CO traces the less collimated outflow cavity provides support that this explanation is plausible.  
The CO data indicate that the outflow is in the plane of the sky, but a slight inclination is necessary to explain the fact that we detect blueshifted SiO emission predominantly to the northwest and redshifted emission predominantly to the southeast indicating that the jet component is not exactly in the plane of the sky.  
In either case, the SiO data are consistent with the CO outflow to the extent that any disk component, if present, should have a position angle roughly 45$^{\circ}$ east of north.


\begin{table*}
\caption{Outflow parameters without inclination angle correction and corrected for an inclination of 20$^{\circ}$ with respect to the plane of the sky.  \label{output}}
\begin{tabular}{ccccccccc}
\hline
\hline
i [$^{\circ}$] & M$_{\rm{t}}$ [M$_{\odot}$] & p [M$_{\odot}$ km/s] & E [erg] & size [pc] & t [yr] & $\dot{\rm{M}}_{\rm{out}}$ [M$_{\odot}$/yr] & F$_{\rm{m}}$ [M$_{\odot}$/km/s/yr] & L$_{\rm{m}}$ [L$_{\odot}$] \\
\hline
  &   13 &   150 & $1.7\times10^{46}$ &  0.45 &  37000 & $3.5\times10^{-4}$ & $4.0\times10^{-3}$ &   3.8 \\
20 &  13 &   440 & $1.5\times10^{47}$ &  0.48 &  14000 & $9.6\times10^{-4}$ & $3.2\times10^{-2}$ &   89 \\
\hline
\end{tabular}
\footnotesize{~\\
Entries include inclination i, total outflow mass M$_{\rm{t}}$, momentum p, energy E, size, outflow dynamical age t, outflow rate $\dot{\rm{M}}_{\rm{out}}$, mechanical force F$_{\rm{m}}$ and mechanical luminosity L$_{\rm{m}}$.  Inputs are discussed in the text.}
\end{table*}

Applying the methods of Cabrit \& Bertout (1990) \nocite{cab90} we derive properties of the system such as outflow mass, dynamical age, and outflow energy. These calculations assume that the \mbox{$^{13}$CO~(2--1)/$^{12}$CO~(2--1)} 
line wing ratio remains constant over the entire outflow.  Based on our single dish data which is consistent with
the data of Choi et al. (1993)\nocite{cho93}, we adopt a value for this ratio of 0.1.
Based on measuring a CO emission spatial extent of 25$\arcsec$ from the central peak and adopting maximum velocities of 10.5 km\ s$^{-1}$ and 13 km\ s$^{-1}$ for the red- and blueshifted lobes respectively, the derived characteristics of the outflow associated with IRDC\ 18223-3 are listed in Table \ref{output}.

The true velocity and spatial extent of an outflow must include a correction for the inclination of the system from the plane of the sky.  Thus any outflow parameters derived from the measured velocity and outflow extent also depend on this correction.  Based on the observation of blueshifted $^{12}$CO emission on both sides of the central source, we interpret this to mean that the outflow is roughly in the plane of the sky.  
In this interpretation, our measurement of the CO spatial extent is a reasonable approximation of the true outflow extent and the effect of inclination on the calculated outflow properties should be small.  The velocity correction, on the other hand, may be significant for inclinations close to 0$^{\circ}$, leading to a much shorter timescale than that listed in Table \ref{output}.  In turn, the outflow rate, mechanical force and mechanical luminosity may be significantly larger than the values given in Table \ref{output} since these values depend on the outflow timescale. 


We compare the quantities presented in Table \ref{output} with the single dish study of massive outflows in Beuther et al. (2002b)\nocite{beu02b}.  Our calculated mechanical force (F$_{\rm{m}}$), outflow mass (M$_{\rm{t}}$), outflow rate ($\dot{\rm{M}}_{\rm{out}}$) and luminosity (L$_{\rm{m}}$) place IRDC 18223-3 well within the massive regime of Figures 4a (F$_{\rm{m}}$ vs. M$_{\rm{core}}$) and 7 (M$_{\rm{t}}$ vs. M$_{\rm{core}}$) and at the lower end of the high-mass regime in Figures 4b (F$_{\rm{m}}$ vs. L$_{\rm{m}}$) and 5 ($\dot{\rm{M}}_{\rm{out}}$ vs. L$_{\rm{m}}$) of Beuther et al. (2002b)\nocite{beu02b}.  In all of these figures, IRDC 18223-3 is well above the low mass sources from the studies by Cabrit \& Bertout (1992) \nocite{cab92} and Bontemps et al. (1996)\nocite{bon96}.  Both of the figures in which IRDC 18223-3 is situated near the lower end of the high-mass regime are luminosity dependent.  The position of IRDC 18223-3 in these plots is a reflection of the source's youth and we expect that the luminosity will increase with time eventually shifting the source well into the regions occupied by the high-mass outflows in Figures 4b and 5.  This general agreement with the properties of the high-mass sources further supports the interpretation that this outflow is associated with a source in the massive regime.  The outflow parameters derived here confirm previous suggestions (\cite{beu05b}, \cite{beu07}) that this truly is a massive star in the making.



It is also interesting to note that in the SiO integrated map in Figure \ref{pdbi}, there is an emission  peak approximately 7$''$ west of the main peak and Figure \ref{outflow_sio} demonstrates that this includes both red- and blueshifted velocities.  We also detect a collimated redshifted CO feature situated 
westward of the primary dust continuum peak.  Puzzlingly, we do not detect a corresponding blueshifted counterpart, nor a millimeter source associated with this CO feature, but it may have a connection to the SiO detection as it appears to originate close to where SiO has a secondary peak.  However, this is 
quite speculative, and we refrain from further comment and interpretation of this collimated redshifted CO emission. 


\subsection{Age of the System} \label{age}
Compared to other massive star formation regions, IRDC\ 18223-3 is extremely young as evidenced by the molecular properties described in Beuther et al. (2005c)\nocite{beu05c}.  The C$^{18}$O line width measured at the main continuum peak of the SMA data is 1.7 km\ s$^{-1}$ whereas it is typically a factor of two to five times larger for molecules that trace similar densities in more evolved massive star formation regions (see e.g. \cite{bro96}, \cite{hat98}, \cite{hof00}, \cite{beu02a}).  In the two secondary continuum peaks, the C$^{18}$O line width is smaller still.  The C$^{18}$O line width at the peak to the northeast is 0.94 km\ s$^{-1}$ and in the southwest peak it is 0.60 km\ s$^{-1}$.  While the line width at the center is less than 2 km\ s$^{-1}$, it is still larger than the line widths in the secondary peaks confirming that the outflow seen in $^{12}$CO(2--1) originates from the primary peak.  The quite narrow central line width that we measure, along with the paucity of chemical species present toward the center, support the interpretation that IRDC 18223-3 is at a very young evolutionary stage.

When the luminosity published by Beuther \& Steinacker (2007)\nocite{beu07} is considered in the framework of the models by Krumholz et al. (2007)\nocite{kru07}, estimates for the age of the system are less than $\sim$10,000 years.  Here we estimate the outflow dynamical age of IRDC 18223-3 to be approximately 37,000 years.  

This timescale calculation depends on the tangent of the outflow's inclination angle 
(with 0$^{\circ}$ being in the plane of the sky). 
Applying a correction for inclination could potentially result in a much younger age than the 37,000 years which we calculate here.  As seen in Table \ref{output} for example, with an inclination angle of 20$^{\circ}$ with respect to the plane of the sky, the age estimate decreases by a factor of nearly 3.  

Both the age estimate based on luminosity and the outflow dynamical age are subject to large uncertainties making it difficult to pinpoint the true age. 
We are rather in support of interpreting the source simply as very young based on the non-detection at infrared wavelengths as well as its not very evolved chemical state.



\subsection{Rotational Structure}

In this section we present maps of the velocity distribution weighted by intensity of the molecules in our SMA data set that are associated with tracing the physical conditions in dense gas.

\begin{figure*}
\includegraphics[angle=-90,scale=.7]{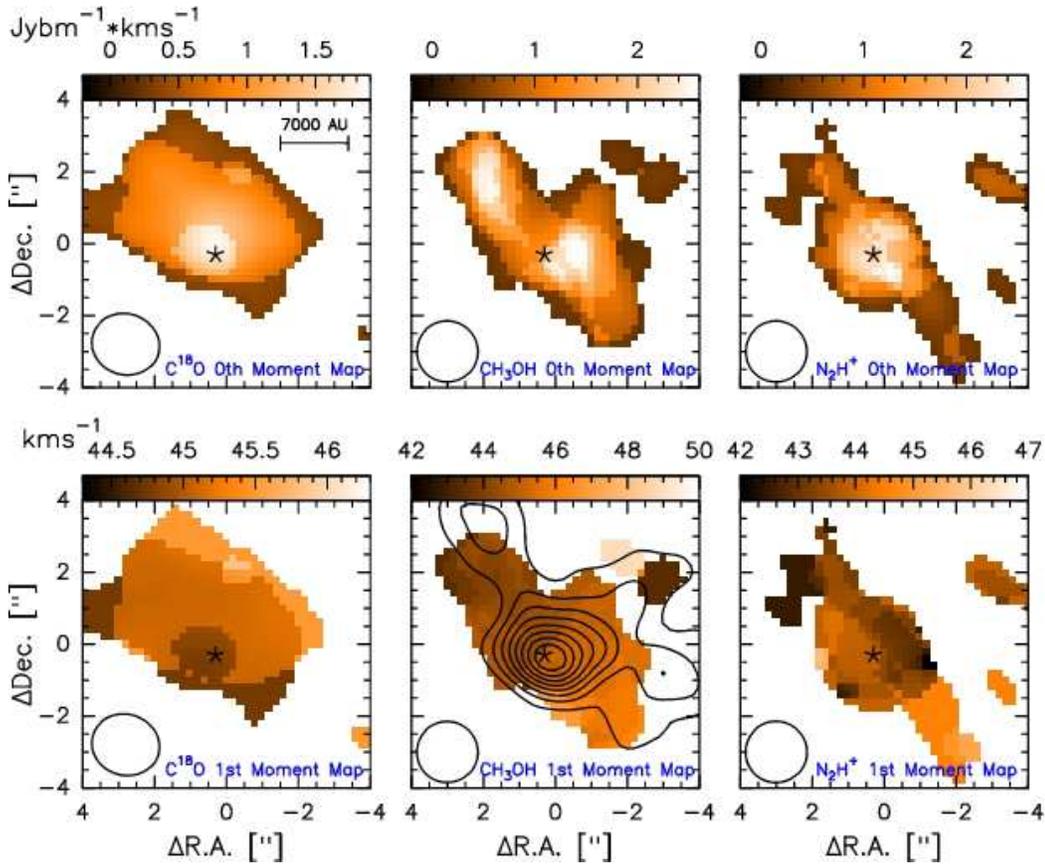}
\caption{\textbf{Upper Row:} C$^{18}$O(2--1), CH$_3$OH(6$_0$-5$_0$) A, and N$_2$H$^+$(3--2) total intensity (0th) moment maps. \textbf{Lower row:}C$^{18}$O(2--1), CH$_3$OH(6$_0$-5$_0$) A, and N$_2$H$^+$(3--2) velocity (1st) moment maps.  The high velocity component toward the northwest corner of the CH$_3$OH velocity moment map is also seen as an extension in the 1.1 mm dust continuum overlaid on the middle panel. The contours start at 5$\sigma$ and increase in steps of 5$\sigma$ where $\sigma$ is 2.25 mJy\ bm$^{-1}$.  All moment maps were clipped at the six sigma level of the respective line's intensity map. The star symbol indicates the position of the 1.3 mm dust continuum peak.  The beam size is shown in the lower left corner of each plot. \label{mom_maps}}
\end{figure*}

The velocity moment map of C$^{18}$O (Figure~\ref{mom_maps}) exhibits a velocity gradient of approximately 2 km s$^{-1}$ from south to north.  
The observed gradient is neither aligned with nor orthogonal to the outflow orientation, presumably because this tracer of denser regions is being strongly affected by infall and the outflow itself. The position angle of the observed gradient differs by about 30$^{\circ}$ with respect to the $^{12}$CO outflow orientation. 
As C$^{18}$O is an isotopologue of $^{12}$CO, it may indeed be tracing a disk-like structure although the observed gradient may be influenced by infall and the outflow.  This is not an entirely unexpected result despite the fact that C$^{18}$O is optically thin and not a good outflow tracer.



Methanol is well documented as a molecule that traces cores, shocks, and masers in star formation regions (e.g., \cite{jor04}; \cite{beu05a}; \cite{sol07}). It has also been reported as a lowmass disk tracer (e.g. \cite{gol99}), yet it has not previously been associated with tracing disks in massive star formation.  In the velocity moment map of the CH$_3$OH(6$_0$-5$_0$) line at 290.110 GHz (middle panel of Figure~\ref{mom_maps}), we see a flattened, elongated rotating structure perpendicular to the outflow axis. The structure evident here is consistent with the velocity moment maps of the next two strongest methanol lines, 229.759 GHz and 290.070 GHz.  The remaining methanol lines in our spectral setup (listed in Table \ref{obs_lines}) are too weak to be able to justify similar statements. The approximately 7.5$''$ extent corresponds to a major axis extent of 28,000 AU at the given distance of 3.7 kpc.  Although this is rather large to be a disk, the narrow velocity range of 3 km s$^{-1}$ coupled with the large spatial extent perpendicular to the outflow support an interpretation that this is a large rotating and infalling core similar to the toroids described in the recent review by Cesaroni et al. (2007).  There is also a high velocity component in the northwest corner of the CH$_3$OH velocity moment map which coincides with the outflow and is likely associated with the extension seen in the 1.1 mm dust continuum overlay of Figure~\ref{mom_maps}.

\begin{figure*}
\subfigure{
\includegraphics[angle=-90,scale=.5]{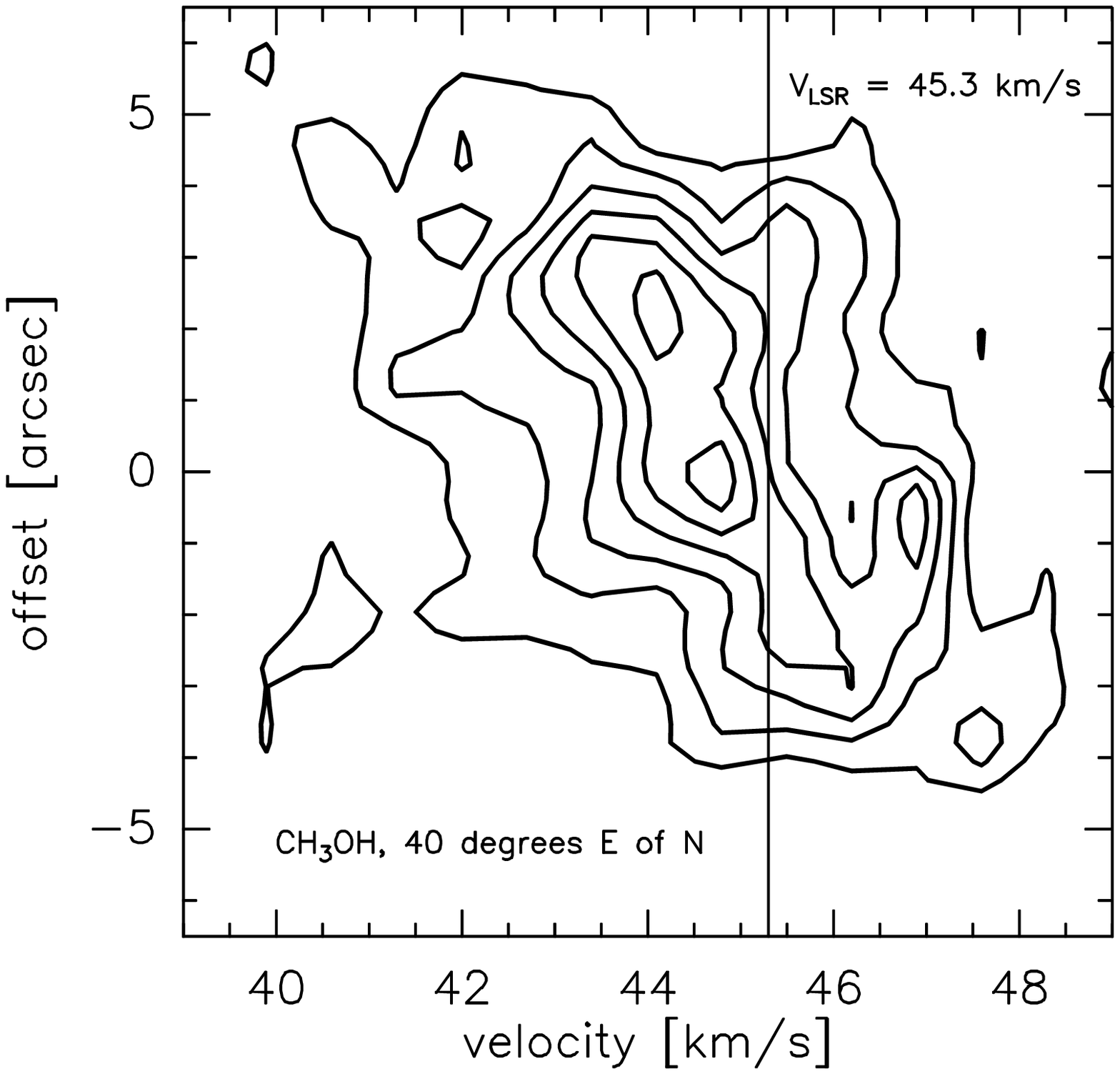}} 
\subfigure{
\includegraphics[angle=-90,scale=.5]{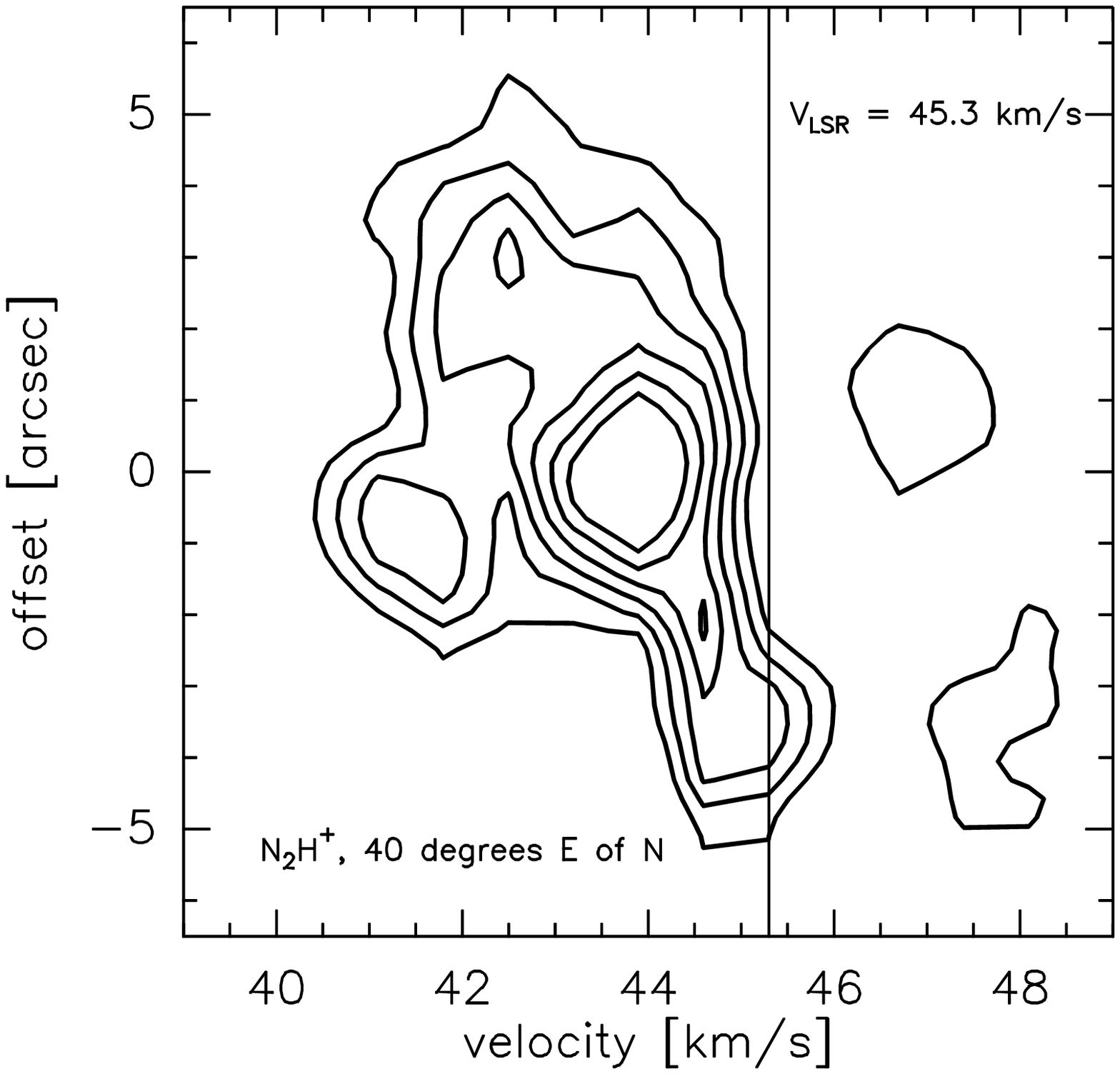}}
\caption{Position-velocity diagrams derived for cuts along the observed velocity gradient.  The offset refers to the distance along the cut from the dust continuum peak.  The rest velocity of 45.3 km\ s$^{-1}$ is marked in both plots.  Contours for both diagrams start at 0.16 Jy and increase in 0.16 Jy steps. \label{pvdiagram}}
\end{figure*}

In Figure~\ref{pvdiagram}, we present the position-velocity diagrams of the CH$_3$OH and N$_2$H$^+$  emission.  The cuts go through the peak of the dust continuum and have position angles along the velocity gradients seen in the moment maps.  In the N$_2$H$^+$ position-velocity diagram, it is apparent that we suffer from missing flux especially around the velocity of rest and at more redshifted velocities, but we still see an indication for the same trends as in the CH$_3$OH position-velocity diagram.  Because of the larger effect of missing flux in N$_2$H$^+$ compared to CH$_3$OH (see Figure~\ref{pvdiagram}), we do not further analyze this molecule suited for probing early stages of star formation, but rather concentrate on CH$_3$OH.

Although the position-velocity diagrams do not exhibit Keplerian velocity signatures, they are consistent with the velocity moment maps in the sense that we see the largest velocities furthest from the center.  
In Keplerian motion, 
the equilibrium condition between gravitational and centrifugal forces means that the velocities at large distances from the source should be lower than those closer to the center. 
One explanation for the deviation from Keplerian motion is that we do not see all the way to the center due to optically thick emission, and hence see only rotation of the outer envelope.  However, without optical depth information, we cannot assume this to be the case, so we interpret the observed velocity gradient simply as non-Keplerian rotation of a large structure perpendicular to the outflow orientation.  This should not come as a surprise as the mass of the central source is likely much less than the mass of the rotating structure.

\subsection{Theoretical Modeling} \label{modeling}

Assuming optically thin line emission, we modeled the velocity gradient seen in the CH$_3$OH velocity moment map.  The modeling was done using the Ulrich Model (\cite{ulr76}, \cite{ket07}) which is an angular momentum conserving mass inflow model that includes gravity and an isothermal rotating envelope.  We calculated the velocity and density along the line of sight and then multiplied these together to produce the velocity distribution.  The model does not include a disk because this adds another parameter which we cannot distinguish between due to observational limitations.  The input parameters we used were determined from the CH$_3$OH moment map.  We measure an observed rotation speed of 1.5 km\ s$^{-1}$ at 14,000 AU (3.8$''$ at the given distance of 3.7kpc).  We chose a mass infall rate of 10$^{-3}$ M$_{\odot}$ yr$^{-1}$.
The system in our current studies does not fit well into the picture of Keplerian rotation that describes low-mass accretion disks (\cite{dut07}), but in order to get a rough estimate of the mass in the rotating structure 
we make the assumption that rotational and gravitational forces are balanced at the outer edge of the disk. We then calculate the mass that is in gravitationally bound rotational motion within a given radius, $r$, using the following equation:
  \begin{eqnarray}
      M_{rot} &=& \frac{\delta v^{2}r}{G} \\
      M_{rot}\mathrm{[M_{\odot}]} &=& 1.13 \cdot 10^{-3} \times \delta v^{2}\mathrm{[km/s]} \times r\mathrm{[AU]}
  \end{eqnarray}
We take $r$ to be 14,000 AU, and $\delta v$ to be 1.5 km\ s$^{-1}$, half the extent of emission and half the velocity range seen in the CH$_3$OH velocity moment map.  From these values, we estimate $M_{rot}$ to be 36 M$_{\odot}$.  
The rather low luminosity of this source leads to the interpretation that this mass is strongly dominated by a larger rotational structure opposed to a central object.
In addition to the large uncertainties (typically within a factor of 5) associated with mass estimates in general, the discrepancy between the mass estimate of 120 M$_{\odot}$ from the continuum emission (Section \ref{mmcontinuum}) and the mass calculated here can be explained by the fact that the system is not rotationally supported but rather infalling, as we assume for the modeling. 

\begin{figure}
\subfigure{
\includegraphics[angle=90,scale=.4]{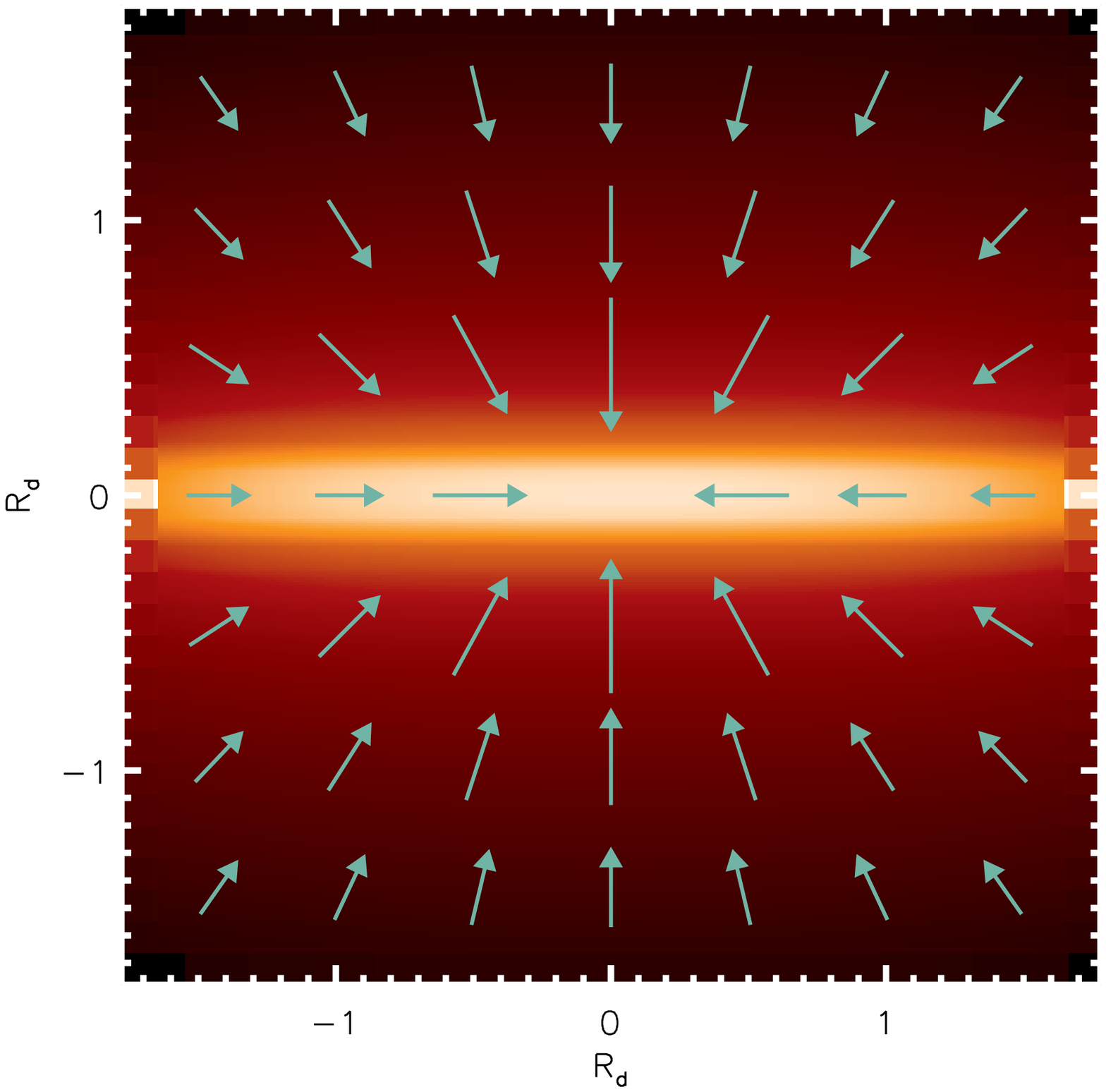}} 
\subfigure{
\includegraphics[angle=-90,scale=.4]{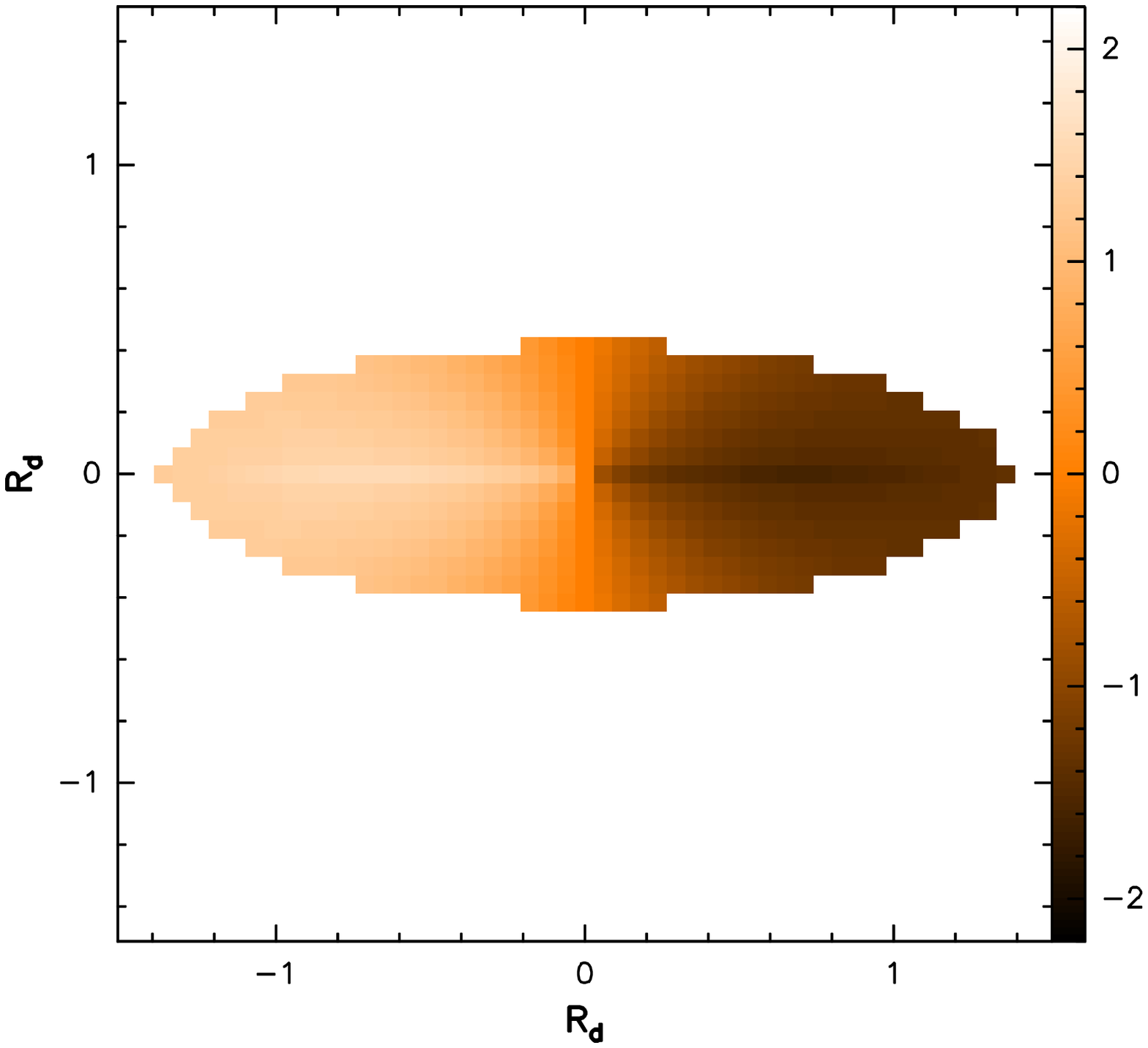}}
\caption{Results of modeling the velocity gradient seen in CH$_3$OH with the Ulrich infall model.  R$_d$ is the radius, and 1 R$_d$ corresponds to 10,000 AU. 
\textbf{Upper:} The density distribution overlaid with the velocity field vectors.  The velocity vectors vary between 0.5 and 1.0 km s$^{-1}$ in length, and the densities vary between 9.4$\times$10$^5$ and 3.9$\times$10$^7$ cm$^{-3}$.  The logarithmic plotting of the density resembles a disk, but we have not included one in this modeling. 
\textbf{Lower:} The velocity and density distributions above multiplied together then averaged along the line of sight to reproduce the velocity gradient seen in the CH$_3$OH velocity moment map. \label{ch3oh_model}}
\end{figure}

The resulting density distribution and velocity field along with the combination of these into the velocity distribution are shown in Figure \ref{ch3oh_model}.  In these plots the disk formation radius, R$_D$, is the radius in the midplane that has the highest density and corresponds to the point at which the rotational force balances the gravitational force.   At the disk formation radius, the Keplerian velocity is 2.1 km\ s$^{-1}$.  We do not observe Keplerian velocity signatures in the position-velocity diagrams (Figure \ref{pvdiagram}), so we do not expect to measure the Keplerian velocity predicted by the model.  Although we do not exactly fit our quantitative parameters, this simple model of rotationally supported infall reproduces our observations well.  

\begin{figure*}
\includegraphics[angle=-90,scale=.7]{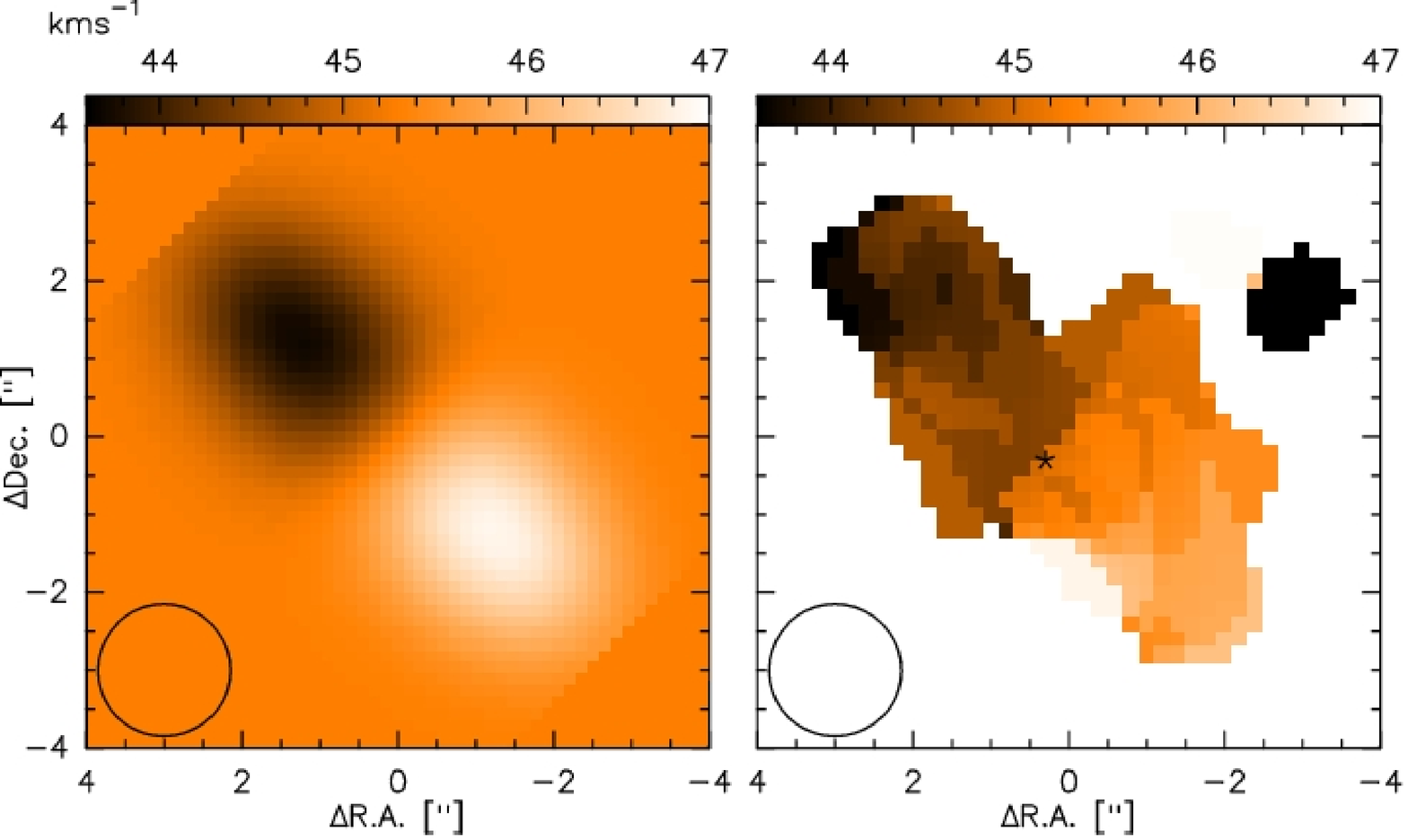}
\caption{\textbf{Left:} The model shown in the lower panel of Figure \ref{ch3oh_model} after convolving to the 1.7$\arcsec$ resolution of the data and rotating counterclockwise by 45$^{\circ}$ into alignment with the observed velocity gradient. \textbf{Right:} The CH$_3$OH velocity moment map as in Figure \ref{mom_maps}.  The velocity range has been modified to exclude the high velocity components seen in the northwest of Figure \ref{mom_maps} so that a more direct comparison to the model can be made. \label{compare}}
\end{figure*}

In order to make a more direct comparison between the theoretical model and our observations, we have convolved the model to the 1.7$\arcsec$ angular resolution of the data and rotated it counterclockwise by 45$^{\circ}$ such that it has the same orientation as the observed velocity gradient.  Figure \ref{compare} presents the smoothed and rotated model and the CH$_3$OH velocity moment map side by side.  We emphasize that this model is not meant to be a best fit of the data, but rather to show that our interpretation presented here is physically plausible.

Keplerian rotation signatures are not detected in this and other regions of massive star formation (see \cite{ces07}). One scenario that could explain this is that as the core collapses, it fragments into smaller condensations and forms several stars, each with accretion disks not necessarily aligned with the parental core's rotational axis.  If this were the case, we would expect to see a more chaotic velocity structure as well as more outflow signatures.  Additionally, this may not be physically reasonable in light of the modeling by Krumholz et al. (2007, 2009). \nocite{kru07,kru09} In these models, multiple stars form within the same accretion disk meaning that the rotational axis should be similar for all sub-sources.  Since a driving source is required to power the outflow, we suspect that a true Keplerian accretion disk may have formed at the center on scales we cannot observationally resolve.

One interpretation of the observed lobes to the northeast and southwest in the millimeter continuum is that there are several unresolved sources in the vicinity.  If these objects are all part of an inward spiraling structure, this might be mistaken for a velocity gradient across them. However, our ability to reproduce the gradient in the CH$_3$OH velocity moment map through successful modeling provides evidence against this interpretation and rather in favor of a scenario in which a single component is surrounded by a large rotating object.  In order to discriminate between these scenarios mentioned here, even higher spatial resolution images of the system would be necessary.  Further studies of spectral lines probing deeper into the core would also likely prove to be helpful.

\section{Conclusions}

The Infrared Dark Cloud 18223-3, a source at the onset of massive star formation, has been observed with multiple submillimeter observatories.  Despite being at a very early stage in the evolutionary process, we detect a molecular outflow and evidence for a large rotating object perpendicular to the outflow.  

This region likely represents one of the earliest stages of disk formation.  IRDC 18223-3 exhibits a large rotating and flattened envelope structure which we suspect is one of the earliest detectable stages in the evolutionary sequence of accretion disks.  In this framework, the structure we see and have described here will continue contracting into a true accretion disk as time passes.

The combined SMA and 30\,m data reveal a well defined northwest-southeast outflow orientation.  To the southeast of the 1.3 mm continuum peak, we see a cone-like blueshifted component and the northwest is dominated by a broader blueshifted component.  
While it is less clear what is happening in the redshifted regime, the blueshifted components indicate that we are likely looking at a system that is roughly in the plane of the sky.  From this we deduce that the likely disk orientation would be approximately edge on and that the associated velocity gradient would be in the northeast-southwest direction.

 Based on the observed outflow properties, we find that calculated parameters are consistent with a massive driving source.  Although the central source is not very massive at this point in time, the outflow characteristics provide a strong indication that this star will evolve into the massive regime.

Indeed we do see velocity gradients in CH$_3$OH and N$_2$H$^+$, although they are observed over a very large spatial distance and a relatively small velocity range.  
On the order of 20,000 AU in size, the large rotating core we currently detect is much larger than other disk or disk-like structures around similar low luminosity intermediate to high mass star forming regions published in the literature.  For instance, the disk associated with IRAS 20126+4104 is less than a third the size of the structure we observe in IRDC 18223-3 (\cite{ces05}).  The reported disks around G192.16-3.82 (\cite{she01}) and IRAS 18089-1732 (\cite{beu04}) are smaller still.
We thus suspect that CH$_3$OH and N$_2$H$^+$ are actually tracing the outer edges of an infalling toroid.  This rotating toroid likely plays an important role in feeding an accretion disk within the unresolved central region  
and may decrease in size over time as the outer edges dissipate or contract leading to growth of the disk.  Hence we see here potentially the earliest stages of the disk formation process.
As there is not much known about CH$_3$OH as a tracer of massive disks, one possibility is that it may be a good tracer for disk kinematics only at very early evolutionary stages, but that other hot core molecules such as CH$_3$CN, for example, may be better for more evolved massive star formation regions.

Using the Ulrich Model for mass infall we have successfully reproduced the velocity gradient observed in methanol.  This agreement between model and observations further enhances the argument that the rotation we see results from a single flattened rotating entity rather than multiple sources or other complicated scenarios.

Similarly, we detect a velocity gradient in C$^{18}$O that is not aligned with the $^{12}$CO outflow axis.  However, the spatial scale is large, and the velocity range is even smaller than in CH$_3$OH and N$_2$H$^+$.  In this case, since C$^{18}$O is an isotopologue of the standard outflow tracer, $^{12}$CO, it is likely that the velocity gradient we see is being influenced by infall motion or the outflow itself 
which would tend to shift the observed gradient toward the direction of the outflow orientation.

The results we have presented further support the idea that IRDCs are sources undergoing the early stages of high-mass star formation.  Thus IRDCs provide ideal sites to study the birth of massive stars.


%
%



\begin{acknowledgements}
C.F. and H.B. acknowledge support by the {\it Deutsche Forschungsgemeinschaft}, DFG project number BE 2578.  C.F. also acknowledges support from the {\it International Max-Planck Research School for Astronomy \& Cosmic Physics} at the University of Heidelberg and thanks Jan Martin Winters for his guidance in reducing the PdBI data sets.  We express our gratitude to the anonymous referee for his/her constructive comments which improved the paper.
\end{acknowledgements}


\begin{thebibliography}{}

\bibitem[Beuther et al.\ 2002a]{beu02a} Beuther, H., Schilke, P., Menten, K. M., et al. 2002a, \apj, 566, 945

\bibitem[Beuther et al.\ 2002b]{beu02b} Beuther, H., Schilke, P., Sridharan, T.K., et al. 2002b, A\&A, 383, 892
\bibitem[Beuther et al.\ 2004]{beu04} Beuther, H., Hunter, T. R., Zhang, Q., et al. 2004, \apj, 616, L39

\bibitem[Beuther et al.\ 2005a]{beu05a} Beuther, H., Zhang, Q., Greenhill, L. J., et al. 2005a, \apj, 632, 355

\bibitem[Beuther et al.\ 2005b]{beu05b} Beuther, H., Schilke, P., Menten, K. M., et al. 2005b, \apj, 633, 535 

\bibitem[Beuther et al.\ 2005c]{beu05c} Beuther, H., Sridharan, T.K., \& Saito, M. 2005c, \apj, 634, L185

\bibitem[Beuther et al.\ 2007]{beu07pp5} Beuther, H., Churchwell, E. B., McKee, C. F., \& Tan, J. C. 2007, in Protostars and Planets V, ed. B. Reipurth, D. Jewitt, \& K. Keil (Tucson: Univ. Arizona Press), 165

\bibitem[Beuther \& Steinacker\ 2007]{beu07} Beuther, H., \& Steinacker, J. 2007, \apj, 656L, 85

\bibitem[Beuther \& Walsh\ 2008]{beu08} Beuther, H., \& Walsh, A. J. 2008, \apj, 673L, 55

\bibitem[Bontemps et al.\ 1996]{bon96} Bontemps, S., Andre, P., Terebey, S., \& Cabrit, S. 1996, A\&A, 311, 858

\bibitem[Bronfman et al.\ 1996]{bro96} Bronfman, L., Nyman, L.-\AA, \& May, J. 1996, A\&AS, 115, 81

\bibitem[Cabrit \& Bertout 1990]{cab90} Cabrit, S., \& Bertout, C. 1990, \apj, 348, 530

\bibitem[Cabrit \& Bertout 1992]{cab92} Cabrit, S., \& Bertout, C. 1992, A\&A, 261, 276

\bibitem[Cesaroni et al.\ 1997]{ces97} Cesaroni, R., Felli, M., Testi, L., Walmsley, C. M., \& Olmi, L. 1997, A\&A, 325, 725

\bibitem[Cesaroni et al.\ 2005]{ces05} Cesaroni, R., Neri, R., Olmi, L., et al. 2005, A\&A, 434, 1039

\bibitem[Ceseroni et al.\ 2007]{ces07} Cesaroni, R., Galli, D., Lodato, G., Walmsley, C. M., \& Zhang, Q. 2007, in Protostars \& Planets V, ed. B. Reipurth, D. Jewitt, \& K. Keil (Tucson: Univ. of Arizona Press), 197

\bibitem[Choi et al. 1993]{cho93} Choi, M., Evans, N. J., \& Jaffe, D. T. 1993, \apj, 417, 624

\bibitem[Dutrey et al.\ 2007]{dut07} Dutrey, A., Guilloteau, S., \& Ho, P. 2007, in Protostars \& Planets V, ed. B. Reipurth, D. Jewitt, \& K. Keil (Tucson: Univ. of Arizona Press), 495

\bibitem[Frerking et al.\ 1982]{fre82} Frerking, M. A., Langer, W. D., \& Wilson, R. W. 1982, 
\apj, 262, 590

\bibitem[Goldsmith et al.\ 1999]{gol99} Goldsmith, P., Langer, W., \& Velusamy, T. 1999, \apj, 519, L173

\bibitem[Hatchell et al.\ 1998]{hat98}Hatchell, J., Thompson, M. A., Millar, T. J., \& MacDonald, G. H. 1998, A\&AS, 133, 29
 
\bibitem[Hildebrand 1983]{hil83} Hildebrand, R. H. 1983, QJRAS, 24, 267

\bibitem[Hofner et al.\ 2000]{hof00} Hofner, P., Wyrowski, F., Walmsley, C. M., \& Churchwell, E. 2000, \apj, 536, 393

\bibitem[J$\o$rgensen et al.\ 2004]{jor04} J$\o$rgensen, J. K., Hogerheijde, M. R., van Dishoeck, E. F., et al. 2004, A\&A, 415, 1021

\bibitem[Keto 2007]{ket07} Keto, E. 2007, \apj, 666, 976

\bibitem[Krumholz et al.\ 2007]{kru07} Krumholz, M., Klein, R., \& McKee, C. 2007, \apj, 656, 959

\bibitem[Krumholz et al.\ 2009]{kru09} Krumholz, M., Klein, R., McKee, C., Offner, S., \& Cunningham, A. 2009, Science, 323, 754

\bibitem[Schilke et al.\ 1997]{sch97} Schilke, P., Walmsley, C. M., Pineau des Forets, G., \& Flower, D. R. 1997, A\&A, 321, 293

\bibitem[Shepherd \& Churchwell 1996]{she96} Shepherd, D. S., \& Churchwell, E. 1996, \apj, 457, 267

\bibitem[Shepherd et al.\ 2001]{she01} Shepherd, D. S., Claussen, M. J., \& Kurtz, S. E. 2001, Sci, 292, 1513

\bibitem[Shepherd et al.\ 2002]{she02} Shepherd, D. S., Claussen, M. J., \& Kurtz, S. E. 2002, ASPC, 267, 415

\bibitem[Sobolev et al.\ 2007]{sol07} Sobolev, A. M., Cragg, D. M., Ellingsen, S. P., et al. 2007, IAUS, 242, 81

\bibitem[Sridharan et al.\ 2005]{sri05} Sridharan, T.K., Beuther, H., Saito, M., Wyrowski, F., \& Schilke, P. 2005, \apj, 634, L57

\bibitem[Ulrich 1976]{ulr76} Ulrich, R. 1976, \apj, 210, 377

\bibitem[Yorke \& Sonnhalter\ 2002]{yor02} Yorke, H., \& Sonnhalter, C. 2002, \apj, 569, 846

\bibitem[Zhang et al.\ 2001]{zha01} Zhang, Q., Hunter, T. R., Brand, J., et al. 2001, \apj, 552, L167

\bibitem[Zhang et al.\ 2005]{zha05} Zhang, Q., Hunter, T. R., Brand, J., et al. 2005, \apj, 625, 864

\bibitem[Zinnecker \& Yorke 2007]{zin07} Zinnecker, H., \& Yorke, H. 2007, ARA\&A, 45, 481

\end{thebibliography}
\end{document}